\documentclass{iopart} 
\usepackage{iopams}  
\usepackage{graphicx,graphics}	
\graphicspath{{images/},{figures/}}
\usepackage[utf8]{inputenc}
\usepackage{eqnarray}
\expandafter\let\csname equation*\endcsname\relax 
\expandafter\let\csname endequation*\endcsname\relax
\usepackage{amsmath,amssymb,bbm} 

\def\notext#1{}

\newcommand{\mcH}{\mathcal{H}}	
\newcommand{\env}{\text{env}}	
\renewcommand{\env}{{\rm env}}	
\newcommand{\osc}{{\rm osc}}

\newcommand{\<}{\langle}
\renewcommand{\>}{\rangle}
\def\1{{\mathchoice{\rm 1\mskip-4mu l}{\rm 1\mskip-4mu l}{\rm 1\mskip-4.5mu l}{\rm 1\mskip-5mu l} }}

\newcommand{\h}{\mathrm{h}}
\newcommand{\qh}{\mathrm{qh}}
\newcommand{\TDJ}{\mathrm{T}_{\Delta,J}}

\newcommand{\hc}{{\rm h. c.}}

\usepackage{hyperref}

\providecommand{\openone}{\leavevmode\hbox{\small1\kern-3.8pt\normalsize1}}

\begin{document} 
\title{Multipartite entanglement dynamics in a cavity} 
\author{J G Amaro$^1$, C Pineda$^2$}
\address{$^1$Facultad de Ciencias, Universidad Nacional Autónoma de México, México D.F. 01000, México}
\address{$^2$Instituto de Física, Universidad Nacional Autónoma de México, México D.F. 01000, México}
\date{\today}%
\begin{abstract}
We study the dynamics of two kinds of entanglement, and there interplay.  On one hand, the intrinsic entanglement within a central system composed by three two level atoms, and measured by multipartite concurrence; on the other, the entanglement between the central system and a cavity, acting as an environment, and measured with purity.  Using dipole-dipole and Ising interactions between atoms we propose two Hamiltonians, a homogeneous and a quasi-homogeneous one.  We find an upper bound for concurrence as a function of purity, associated to the evolution of the $W$ state. A lower bound is also observed for the homogeneous case. In both situations, we show the existence of critical values of the interaction, for which the dynamics of entanglement seem complex. 
\end{abstract}
\pacs{03.65.Yz,03.67.-a}
\section{Introduction}  

Entanglement was appreciated by Schrödinger not as ``{\it one} but rather {\it
the} characteristic trait of quantum mechanics''~\cite{Schrodinger1935}.  Its
study has opened its own field field~\cite{RevModPhys.81.865,
RevModPhys.84.1765}, that has remain active because the understanding of
entanglement has
proved valuable to comprehend the transition from the quantum to the classical
world. Also, from a pragmatic point of view, to build ever more
complex quantum technologies, one needs to tame its effects.  Entanglement has
been identified as a key resource in, say, quantum simulation and quantum
computing~\cite{RevModPhys.84.1655}. 

Several theoretical developments have been devoted to understand the evolution
of the entanglement of bipartite systems, both from theoretical, and an
experimental point of view (see for example,~\cite{citeulike:2365852,
PhysRevA.73.012305, PhysRevA.82.042328} and many articles citing these ones).
Nowadays, two party entanglement is routinely produced, controlled, studied and
exploited in the laboratory.  Moreover, it is well understood. Many aspects of
multipartite entanglement, despite a huge effort, still remain as open
questions (see for example~\cite{7443} and \cite{pregunta}).  In this paper we
shall study the simplest case of multipartite entanglement, namely three
qubits, and its interplay with bipartite entanglement, often associated with
decoherence. For this, we use the simplest possible ``reservoir'' with an
infinite spectrum: a harmonic oscillator.

Building upon~\cite{1751-8121-43-19-192002} we generalize there model, to allow
for more qubits, but retaining its simplicity.  We consider three two-level
atoms, coupled to each other via dipole and Ising interaction. Additionally, we
consider a cavity which shall be regarded as an {\it environment}, to which the
atoms are coupled to a single mode. Both the qubits and the cavity have its
free dynamics.  Except for very special cases, this model is not analytically
solvable. However, number conservation will allow us to work on a finite
Hilbert space, and perform numerics. We shall focus on the evolution of the
entanglement withing the three qubits, and of those, with the cavity, as a
measure of decoherence. More over, we shall also study the relation between
these two very different quantities, via a generalized $CP$  (concurrence-purity)
map~\cite{ziman:052325,PhysRevA.73.012305,1751-8121-43-19-192002}. 


Our paper is organized as follows. In section \sref{sec:model} we 
introduce the model, and discuss some of its symmetries. We devote
section \sref{sec:entanglement} to recall some aspects of entanglement
that shall be discussed here. Special attention will be given to
multipartite mixed entanglement, as quantified by the {\it concurrence}.
In \sref{sec:results} we present our findings, and we finish 
with some conclusion in \sref{sec:conclusions}. 
\section{The model}  
\label{sec:model}
We shall work in the Hilbert space associated to a harmonic oscillator, and
three qubits. That is $\mcH = \mcH_{\rm ho} \otimes \mcH_{\rm q, 1}\otimes
\mcH_{\rm q, 2}
\otimes \mcH_{\rm q, 3}$ with  $\mcH_{\rm ho}$ the Hilbert space of the 
oscillator, and the others, the Hilbert space of each of the three qubits ($\dim \mcH_{\rm q, i} = 2$).
The Hamiltonian that will determine the evolution of the system is 
\begin{eqnarray}
\fl
\hat H = 
\sum_{j=1}^{3} \frac{\Delta_j}{2} \hat \sigma_z^{(j)} + 
\sum_{j=1}^{3} g_j (\hat a \hat \sigma_+^{(j)} + \hc ) \nonumber \\
+2 \sum_{j \ne k=1}^{3} \kappa_{jk}( \hat \sigma_-^{(j)} \hat \sigma_+^{(k)} + \hc)
 +\sum_{j \ne k=1}^{3} J_{jk} \hat \sigma_z^{(j)} \hat \sigma_z^{(k)},
\label{eq:hamiltonian}
\end{eqnarray}
in which we are using Pauli matrices $\hat \sigma_{x,y,z}^{(j)}$ acting in the
$j$th
spin $1/2$ particle, the lowering and rising
operators of an harmonic oscillator ($\hat a$, $\hat a^\dagger$) and the rising
and lowering
operators of spin $1/2$ particle $j$, $\hat \sigma_\pm^{(j)} = \hat
\sigma_x^{(j)} \pm \rmi \hat \sigma_y^{(j)}$. 
We are also introducing several parameters, namely the energy splitting 
in each qubit ($\Delta_j$), the intensity of their interaction with 
the harmonic oscillator $g_j$, the pairwise dipole-dipole interaction 
$\kappa_{jk}$ and  $J_{jk}$ the Ising interaction. 
Intrinsic dynamics in the oscillator can be safely ignored using the appropriate
interaction picture. 

Note that despite the simplicity of the model, it will admit several very
distinct dynamical configurations, ranging from: all qubits interacting with
each other; a ``line'' configuration in which there is no interaction between
qubits 1 and 3; a spectator configuration in which one qubit does not interact
with the other qubits; and finally one in which all qubits are decoupled from
each other. This can be controlled by setting to zero several of the
coupling parameters $\kappa_{jk}$ and  $J_{jk}$. \par 
A very important feature, that allows us to treat the model, is that
this Hamiltonian preserves the number of excitations, characterized by the operator 
\begin{equation}
\hat N = \frac{1}{2} \sum_{j=1}^{3} \hat \sigma_z^{(j)}
+ \hat a^\dagger \hat a  + \frac32 \openone.
\label{}
\end{equation}
One can thus write the Hamiltonian in block-diagonal form in a suitable basis. We
take this basis to be 
\begin{eqnarray}\label{eq:basegeneralizada} 
|\phi_{0}^{(n)}\> & = |n\> |000\>, \qquad     \quad  \ \ \ \,       |\phi_{4}^{(n)}\> &= |n-2\> |110\>,\nonumber\\
|\phi_{1}^{(n)}\> & = |n-1\>   |001\>,\quad \quad  \quad  |\phi_{5}^{(n)}\> &= |n-2\> |101\>,\nonumber\\
|\phi_{2}^{(n)}\> & = |n-1\>   |010\>,\quad \quad   \quad |\phi_{6}^{(n)}\> &= |n-2\> |011\>,\nonumber\\
|\phi_{3}^{(n)}\> & = |n-1\>   |100\>,\quad \quad  \quad  |\phi_{7}^{(n)}\> &= |n-3\> |111\>,
\end{eqnarray} 
so that $\hat N |\phi_{i}^{(n)}\> = n |\phi_{i}^{(n)}\>$. We are using the
convention in which 
$|1\>$ means an excitation and $|0\>$ means no excitation, so $\sigma_+|0\> =|1\>$. 
Note that, when using matrix representation, even though their is a one to one 
correspondence with 
the state of the qubits, there is additional information
regarding the state of the harmonic oscillator, via de superindex $(n)$. 
The order chosen for the basis will allow us to write the partial trace in a
particularly nice fashion.  Given the matrix representations of an arbitrary
mixed state $\rho$ acting in the subspace of $n$ excitations, we can write 
\begin{equation} 
\tr_\osc \rho = 
\begin{pmatrix}
\rho_{00} &    0      &    0      &    0      &    0      &    0      &    0      &    0      \\
   0      & \rho_{11} & \rho_{12} & \rho_{13} &    0      &    0      &    0      &    0      \\
   0      & \rho_{21} & \rho_{22} & \rho_{23} &    0      &    0      &    0      &    0      \\
   0      & \rho_{31} & \rho_{32} & \rho_{33} &    0      &    0      &    0      &    0      \\
   0      &    0      &    0      &    0      & \rho_{44} & \rho_{45} & \rho_{46} &    0      \\
   0      &    0      &    0      &    0      & \rho_{54} & \rho_{55} & \rho_{56} &    0      \\
   0      &    0      &    0      &    0      & \rho_{64} & \rho_{65} & \rho_{66} &    0      \\
   0      &    0      &    0      &    0      &    0      &    0      &    0      & \rho_{77} 
\end{pmatrix}.
\label{}
\end{equation} \par 
A generic block of $n \ge 3$ excitations 
given by 
\begin{equation}
H^{(n)} =
\begin{pmatrix}
H_{AA}^{(n)} &   H_{AB}^{(n)} \\
\left(H_{AB}^{(n)}\right)^\dagger & H_{BB}^{(n)} 
\end{pmatrix}-\sum_{j \ne k=1}^{3} J_{jk} \openone
\label{eq:hamiltonian:n}
\end{equation}
with the first block being
\begin{equation}
H_{AA} =
\begin{pmatrix}
2J& g_1^+ & g_2^+  & g_3^+ \\
g_1^+ & \delta_1^+  & \kappa_{12} & \kappa_{13} \\
g_2^+ &  \kappa_{12} &  \delta_2^+   & \kappa_{23} \\
g_3^+  & \kappa_{13} & \kappa_{23} &   \delta_3^+
\end{pmatrix}-\sum_{j=1}^3\Delta_j\openone,
\end{equation}
if $g_i^\pm = g_i \sqrt{n \pm 1}$, $\delta_i^\pm =  2J_{\ne i} \pm 2\Delta_i +$, and
$\TDJ = \sum_i J_{\ne i}-\Delta_i$. Note that we use the subindex $\ne i$ to refer
to a pair of subindices, that are different from $i$ and from each other. 
The off-diagonal matrix is 
\begin{equation}
H_{AB} =
\begin{pmatrix}
0&0&0&0\\
0&g_3^0 &g_2^0 &0  \\
g_3^0 &0  &g_1^0 &0 \\
g_2^0 &g_1^0 &0 &0 
\end{pmatrix}
\label{}
\end{equation}
with $g_i^0 =  g_i \sqrt{n}$. Finally, the second diagonal block is given by 
\begin{equation}
H_{BB} =
\begin{pmatrix}
\delta_3^- & \kappa_{12} & \kappa_{13} & g_1^-\\
\kappa_{12} & \delta_2^- & \kappa_{23} & g_2^-\\
\kappa_{13} & \kappa_{23} & \delta_1^- & g_3^-\\
g_1^-&g_2^-&g_3^-&2J
\end{pmatrix} +\sum_{j=1}^3\Delta_j \openone
\label{}
\end{equation}
%
For smaller values of $n$, the matrix will be the first block of $1$, $4$, and $7$
states corresponding to \eref{eq:hamiltonian:n}, for $n=0,1,2$ respectively. \par
This model has a big number of parameters. A simplification, would be to consider 
a homogeneous situation, that is, all $\kappa_{ij}=\kappa$, 
$J_{ij}=J$, $g_i=g$ and $\Delta_i = \Delta$. Moreover, we shall set the
detuning to be $\Delta = 0$ and the coupling to the oscillator to $g =1$ 
in our numerical calculations, so that we have the {\it homogeneous} Hamiltonian
\begin{equation}
\hat H^{\h}_{\kappa,J} = 
\sum_{j=1}^{3} (\hat a \hat \sigma_+^{(j)} + \hc ) 
+2 \kappa \sum_{j \ne k=1}^{3} ( \hat \sigma_-^{(j)} \hat \sigma_+^{(k)} + \hc)
 + J \sum_{j \ne k=1}^{3} \hat \sigma_z^{(j)} \hat \sigma_z^{(k)}.
\label{eq:hh}
\end{equation}
In order to appreciate the effect of homogeneity in the Hamiltonian, we 
also consider a slightly inhomogeneous case: we change one dipole-dipole 
interaction, to obtain the {\it quasi-homogeneous} Hamiltonian
\begin{equation}
\hat H^{\qh}_{\kappa,J} = 
\hat H^{\h}_{\kappa,J}  + \kappa\left( \hat \sigma_-^{(1)} \hat \sigma_+^{(2)} + \hc \right).
\label{eq:qh}
\end{equation}
Several other simple ways of introducing the inhomogeneity are also available, 
however, this one displays very clearly the effects we want to underline 
in this work.

Note that one further consideration, in the case of the homogeneous case,
could be done: the Hamiltonian is invariant under rotations.  In this case, the
rotation operator $\hat R$ that acts on the computational basis as $\hat R |n\>   |i_1
i_2 i_3\> = |n\>   |i_3 i_1 i_2\>$, $i_j = \pm$, has three eigenvalues, namely
$\exp(2 \pi \rmi j/3)$, with $j=0,1,2$. Let us define the vectors
\begin{align}
|\varphi_{0}^{(n,k)}\> &=
\alpha^k|\phi_{1}^{(n)}\>+\alpha^{2k}|\phi_{2}^{(n)}\>+\alpha^{2k}|\phi_{3}^{(n)}\>,\\
|\varphi_{1}^{(n,k)}\> &=
\alpha^k|\phi_{4}^{(n)}\>+\alpha^{2k}|\phi_{5}^{(n)}\>+\alpha^{2k}|\phi_{6}^{(n)}\>,
\end{align}
($k=0,1,2$) with $\alpha = \exp(2\pi \rmi /3)$ and the additional
$|\varphi_{2}^{(n,0)}\> = |\phi_{0}^{(n)}\>$ and $|\varphi_{3}^{(n,0)}\> =
|\phi_{7}^{(n)}\>$. These are eigenvectors of $\hat R$ with eigenvalues $\alpha^k$,
that is, $\hat R |\varphi_{l}^{(n,k)}\> = \alpha^k |\varphi_{l}^{(n,k)}\>$.  This
lead to a splitting of the subspace in spaces of dimension 4, 2 and 2. Very
lengthy expressions however, make it very difficult to extract the general
behaviour, and one would be forced to fall back to numerics. On the other hand,
if one wishes to restrict to the symmetric subspace, the machinery of the Dicke
states could be used, where some analytical results, regarding entanglement,
are available~\cite{ PhysRevA.78.052105, 1367-2630-11-7-073039}.

%
%
%


\section{Multipartite entanglement}  
\label{sec:entanglement}

The notion of entanglement is defined using separability. 
Separable pure states $|\psi \>$ are those for which, in a multipartite Hilbert
space $\mcH = \otimes_i \mcH_i$, can be written as a tensor product. That is, 
$|\psi \> \in \mcH$ is separable if $|\psi \> = \otimes_i |\psi_i \>$, with 
$|\psi_i \>\in \mcH_i$. Entangled states are those with are not a mixture
of pure separable states. The problem of determining if a mixed state
is or not entangle is difficult. 

However, for the special case of pure bipartite states, all the information 
regarding entanglement is encoded in its Schmidt coefficients. One can 
choose any convex function of the Schmidt coefficients, being the von Neumann
entropy and purity the most common choices. We shall use purity, defined
for mixed states as 
\begin{equation}
P( \rho ) = \tr \rho^2,
\label{}
\end{equation}
due to its algebraic simplicity. 
Purity can be regarded as an entanglement measure, provided that 
the state over which purity is calculate is given by 
\begin{equation}
\rho = \tr_\env |\psi \> \<\psi | .
\label{}
\end{equation}
If no {\it a priori} information is given about $\rho$, purity is
simply a measure of mixedness. The value of purity ranges between 
$1/N \le P \le 1$, where $N$ is the dimension of the Hilbert space
in which $\rho$ acts, the minimum
value corresponding to the maximally mixed state, and the maximum 
to a pure one.

%

Characterizing multipartite entanglement on the other hand proves more
challenging, as even a unique maximally entangled state does
not exist for more than two parties~\cite{PhysRevA.62.062314}. Most measures
provide either well founded physical grounds, or numerically simple recipes. A
convenient 
compromise is given by the multipartite concurrence. This measure, 
a generalization of the two party concurrence~\cite{wootters}, is inspired in the
symmetry properties of pure states. That is, in expected values
of projections over antisymmetric subspaces. The detailed construction 
is out of reach within this presentation, but it reduces to the very simple
form~\cite{Mintert2005207} for pure states:
\begin{equation}
C(|\psi \> ) = \frac{1}{2^{N/2-1}} \sqrt{\left(2^N-2\right) 
 - \sum_i \tr \rho_i^2}
\end{equation}
where the index $i$ runs over all proper subsets of particles, except for the
empty set. For example, in the three particle case, we must consider all tree
partitions in which the particles can be divided, that is particle A against
BC; B with AC; and C with AB.  Thus, $C(|\psi \> )^2 \propto \<  1- P(\rho_i)
\>$ where here, the average $\< \cdot \>$ is taken again over all non-trivial subsets
particles, thus relating the measure with the entanglement over all possible
partitions.

This measure, as presented above, will not suit our purposes, as it is defined only for
pure states. We shall use the {\it convex roof} construction, in which 
the measure is averaged over a particular realization of our mixed state.
Say, our state $\rho = \sum_i p_i |\psi_i \>\< \psi_i|$, with normalized
$|\psi_i \>$ and positive $p_i$ for all $i$. Then, we associate with this
particular realization of $\rho$ the measure $\sum_i p_i C(|\psi_i \> )$. The
measure is obtained finding the realization that minimizes such expression. This
would mean, if one thinks about entanglement as a resource, the cheapest 
way of realizing such ensemble. 
Summarizing, we define 
\begin{equation}
C(\rho ) = \inf_{p_i, |\psi_i \> } 
\sum_i p_i C(|\psi_i \> ), 
\label{eq:definition:c}
\end{equation}
with $\rho = \sum_i p_i |\psi_i \> \<\psi_i|$, $p_i >0$, and $\<\psi_i |\psi_i \>=1$.
Even though this is a very meaningful definition, the process of exact evaluation
is normally difficult, as the landscape can be very complex. Upper bounds can
be easily obtained using gradient methods to numerically optimize \eref{eq:definition:c},
and also simple, but useful lower algebraic bounds are available. 
In particular, we shall use the method of {\it quasi-pure
approximation}~\cite{PhysRevA.72.012336}. It is derived
approximating a multi-matrix representation of a mixed state with a single
matrix that captures the first few order terms in expressions
involving the Schmidt coefficients of $\rho$. 
We underline that this is not only an
approximation, but also a lower bound, reasonably tight even for some states close to the
maximally mixed state~\cite{Mintert2005207, PhysRevLett.98.140505}. To obtain
this bound, no optimization procedure is involved, and only the diagonalization
of a matrix of
the same size as $\rho$ is needed. A detailed
description requires some technicalities, that we do not wish to introduce. Instead, 
we refer the reader to~\cite{Mintert2005207}. 
Some properties of the concurrence should be highlighted: it is invariant under
unitary local operations, it vanishes only for completely
separable states, and has the nice scaling property that $C(|\psi \>\< \psi|
\otimes |\phi \>\< \phi|) = C(|\phi \>\< \phi|)$ if $|\psi \>$ belongs to the
Hilbert space of just one of the particles. Finally, the bound coincides with
the concurrence for pure states.

\begin{figure} 
   \includegraphics{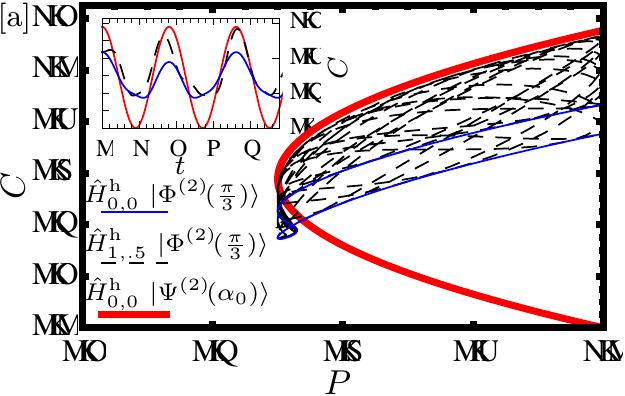} \hfill \includegraphics{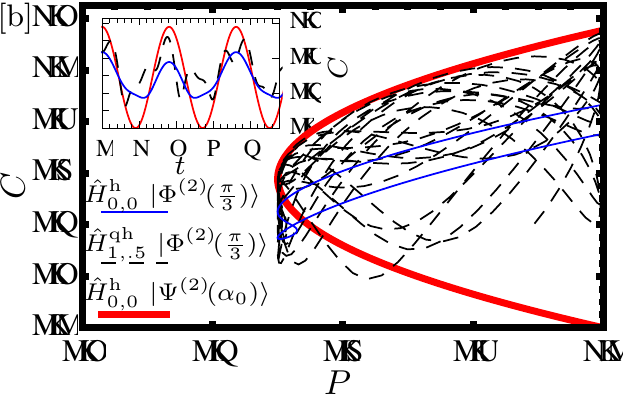}
     \caption{
     Parametric plot of concurrence and purity, for initial states
     of the form~\eref{eq:torres} and \eref{eq:pineda}, with Hamiltonians~\eref{eq:hh} and \eref{eq:qh} 
     and an initial number of excitations of $n=1$. 
     The red curve corresponds to three non-interacting atoms ($\kappa = J =
     0$), initially in a W state.  The black dashed curve shows the behaviour of three
     interacting atoms, with $\kappa = 1$ and $J=0.5$,  initially in a pure but not
     maximally entangled state
     ($\alpha =\pi/3)$. The homogeneous~\eref{eq:hh} and quasi-homogeneous~\eref{eq:qh} configurations
     are displayed in 
     [a], and in [b] respectively. The blue curve shows the evolution of
     the decoupled system, with the same initial state as in the black dashed
     curve.  All curves were parametrized by time up to $t=20$. The insets in
     displays the corresponding evolution of concurrence as function of time.}
         \label{fig:torres_pi4_las_DOS}
\end{figure} 

The quasi-pure approximation suits very well our needs, where repeated
evaluation of such a quantity is required, for many parameters and times. 
Moreover as we shall explore qualitative properties, so the small errors
inherent in this approximation can be ignored.

\section{Results} 
\label{sec:results}
In this  section we calculate the entanglement and purity as a function 
of time, and how they depend on each other. We consider
two families of initial product states with the condition that any member of
the family must belong to the eigenspace characterized by a fixed eigenvalue of
the operator $\hat N$.  \par 
While the first family corresponds to the normalized product state with $n-1$
photons in the cavity and the superposition of the states $|001\>$ and
$|010\>$,
\begin{equation}
|\Phi^{(n)}(\alpha)\>=|n-1\> \otimes \left( \sin \alpha |001\> + \cos \alpha |010\>  \right). 
\label{eq:torres}
\end{equation} 
This states have no genuine tripartite entanglement, as one of the 
parties has a uncorrelated state, and the others share bipartite entanglement parametrized by 
$\alpha$. Its concurrence is 
$C(|\Phi^{(n)}(\alpha)\>) = \sin(2\alpha)$.
The second family corresponds to  the superposition of the states
$|001\>$, $|010\>$ and $|100\>$,
\begin{equation}
|\Psi^{(n)}(\alpha)\>= |n-1\> \otimes
\left( \frac{\sin \alpha}{\sqrt{2}} |001\>+\cos \alpha |010\>+ \frac{\sin \alpha}{\sqrt{2}}|100\>\right).
\label{eq:pineda}
\end{equation}
This states do have genuine tripartite entanglement, as can be seen by setting 
$\alpha=\alpha_0 = \arctan\sqrt{2}$  for which we retrieve the $|W\>$ state. 
Moreover, its concurrence is 
\begin{equation}
C(|\Psi^{(n)}(\alpha)\>) = \frac{\sin \alpha}{\sqrt{2}} \sqrt{5+3\cos (2\alpha)},
\end{equation}
with a maximum given by $C(|\Psi^{(n)}(\alpha_0)\>) =2/\sqrt{3}\approx 1.15 $. 
However, in 
both cases we can range from a totally separable state for $\alpha=0$ to a maximally 
entangled state for some critical $\alpha$.  These families of states also have the
important characteristic that will remain pure when the oscillator 
is traced out. 
\begin{figure}
   \includegraphics{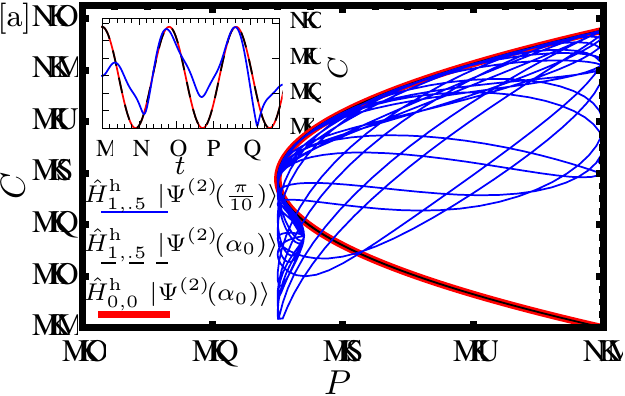} \hfill \includegraphics{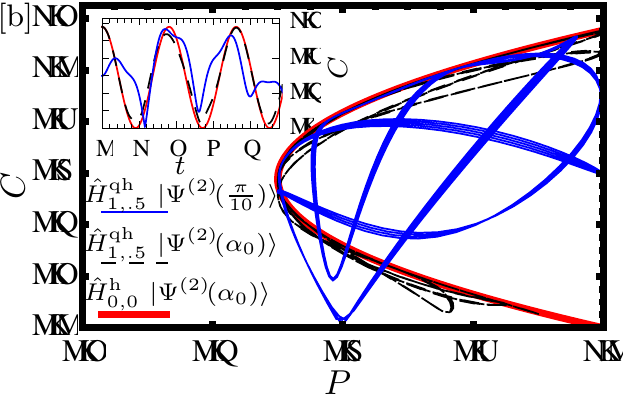}
     \caption{
      Same conventions as in \fref{fig:torres_pi4_las_DOS}, except that the initial states 
      for the interacting case are 
      given by \eref{eq:pineda}. The parameters are identical to the aforementioned figure. 
     }
         \label{fig:carlos_las_DOS}
\end{figure}
The system evolves with the unitary evolution generated by Hamiltonian 
\eref{eq:hamiltonian}, and the state of the three qubits is given at
time $t$ by 
\begin{equation}
\rho(t)
= \tr_{\rm ho} \rme^{-\rmi t \hat H} |\psi(0)\>\<\psi(0)| \rme^{\rmi t \hat H},
\label{}
\end{equation}
where $|\psi(0)\>$ is the initial state of the system, which will be taken as
either $|\Phi^{(n)}(\alpha)\>$ or $|\Psi^{(n)}(\alpha)\>$. One can then study the
evolution of concurrence and purity, by studying $C(t) = C(\rho(t))$ 
and $P(t) = P(\rho(t))$ respectively. Moreover, we shall study an homogeneous
case, in which all $J_{ij} = J$ and all $\kappa_{ij} = \kappa$. \par
In figures~\ref{fig:torres_pi4_las_DOS} and \ref{fig:carlos_las_DOS} we present
concurrence against purity with time as a parameter, that is, the so called
$CP$ plane. We studied in these figures the sector $n=1$, but the conclusions
here drawn can be extended to higher excitation numbers, unless we explicitly
say otherwise. We also show in the inset the evolution of concurrence. We fix
the Hamiltonians to the homogeneous and quasi-homogeneous with parameters
$J=0.5$ and $\kappa=1$. That is, using the notation of \eref{eq:hh} and
\eref{eq:qh}, we use $H^\h_{1,0.5}$ and $H^\qh_{1,0.5}$. In all cases an
important benchmark shall be considered, namely a maximally mixed state
$|\Psi^{(n)}(\alpha_0)\> \propto |001\> + |010\> +  |100\> $, evolved with an
interaction free model, that is, $H^\h_{0,0}$.  This is plotted as a thick red
curve.  \par
In figure~\ref{fig:torres_pi4_las_DOS} we study initially non-maximally
entangled states, both with and without interaction.
In both
figures
\ref{fig:torres_pi4_las_DOS}, one can note that the red curve serves as an
upper bound for the evolution of the states without maximal initial
entanglement, similar to the two atom case~\cite{1751-8121-43-19-192002}.
Remarkably, for three atoms, the dynamics of the non-interacting case
is a lower bound in the homogeneous case [a] (as in the two qubit case), whereas
that is no longer the case when the system is not homogeneous~[b]. 
We underline here that in all cases studied (and not shown), namely other parameters, 
and excitation numbers, these observations hold. 
The behaviour of concurrence and purity with time (displayed in the inset)
will be discussed later, when we have a global picture, with respect to the 
parameters and time. However, it can be noted that the non homogeneous case 
display a richer behaviour. \par
\begin{figure}
   \includegraphics[width=\textwidth]{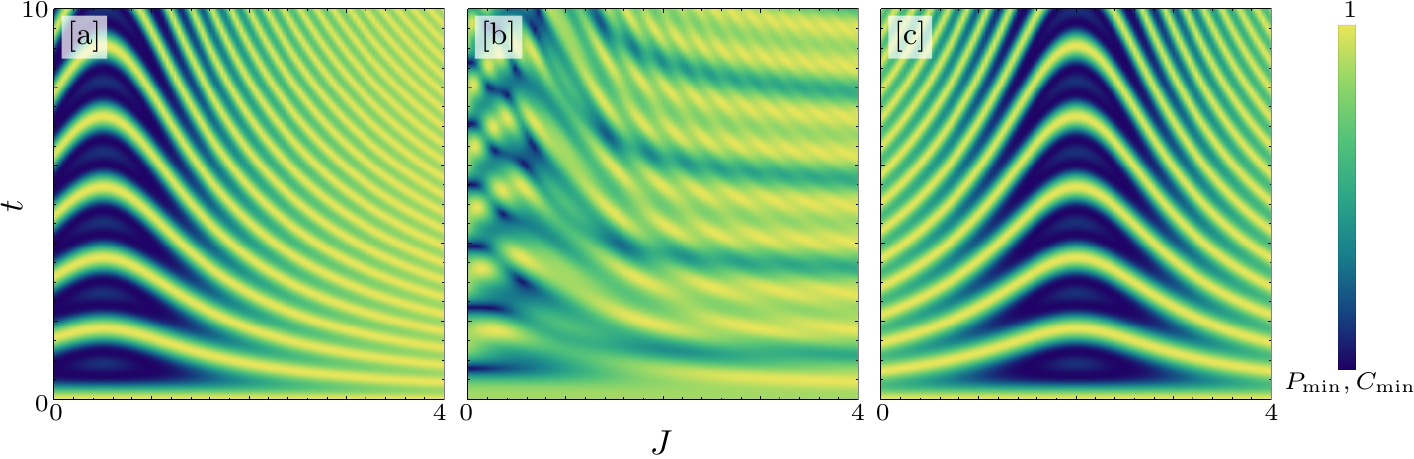}
     \caption{
     In these set of figures, we study the behaviour of purity and concurrence
     (as colour density, in arbitrary units) with the Ising interaction (in the
     $x$ axis) and time (in the $y$ axis) for the initial state
     $|\Phi^{(2)}(\pi/4)\>$ and the Hamiltonian $\hat H^{\h}_{\kappa,J}$.  In
     [a] we show purity, with $\kappa=1$, and observe no complex behaviour.
     Concurrence, plotted in [b], mimics this behaviour of purity, with 
     added complexity. A larger dipole-dipole coupling ($\kappa=4$), shown in
     [c], displaces the bump observed in [a], to larger values of $J$. }
     \label{fig:traslado}
\end{figure}
In \fref{fig:carlos_las_DOS} we study the effect of interactions on 
both maximally entangled initial states $|\Psi^{(n)}(\alpha_0)\>$, 
and partially entangled states $|\Psi^{(n)}(\pi/10)\>$. In this case, 
 the behaviour of states which have initially maximum
entanglement, with already large interaction, 
closely resemble the case with no interaction. In particular, for the homogeneous
Hamiltonian the two curves coincide,  while for the quasi-pure Hamiltonian
the non interacting case acts as an imperfect, but very good guide.
This is not the case if we start with a state with smaller entanglement. 
The red curve acts only as an upper bound
for the interacting case. For larger values of the Ising interaction the
dynamics display an oscillatory
behaviour with almost the same periodicity. These observations
are robust with respect to varying the systems and the total 
number of excitations. \par
\begin{figure}
   \includegraphics[width=\textwidth]{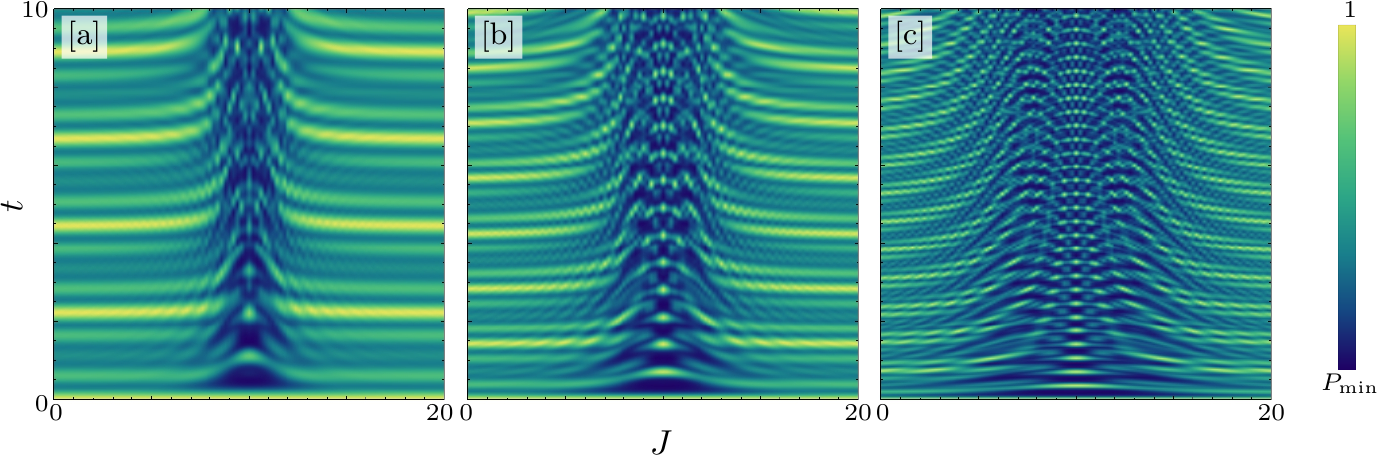}
     \caption{
     In these set of figures, we study the behaviour of purity with the Ising
     interaction (in the $x$ axis) and time (in the $y$ axis), 
     for 
     a varying number of total excitations $|\Phi^{(2,5,20)}(\pi/4)\>$  in [a], [b]
     and [c] respectively, evolved with $\hat
     H^{\h}_{20,J}$.  The region with complex behaviour increases
     with the number of excitations. 
     }
     \label{fig:ensanchmiento}
\end{figure}

In order to give a general view of the dynamics of purity and concurrence we
proposed a density diagram display in figures \ref{fig:traslado} and
\ref{fig:ensanchmiento}. There, concurrence and purity are colour coded for
several values of the Ising interaction, for a given time span,  under the
influence of the homogeneous Hamiltonian. The initial state chosen was
$|\Psi^{(2)}(\alpha=\pi/4)\>$. 

In figure~\ref{fig:traslado} we compare the dynamics of purity
(figures~\ref{fig:traslado}[a,c]) and  concurrence
(figure~\ref{fig:traslado}[b]). Indeed, concurrence mimics the behaviour of
purity, with some additional nodes, caused possibly by the internal dynamics of
the three qubits. 

Purity displays a regular behaviour, with some some oscillations that seem to
be independent of $J$ for large or small values of $J$. However, a critical
area (around $J=0.5$ for $\kappa=1$, figure~\ref{fig:traslado}[a]) is clearly
present.  Varying the dipole-dipole coupling shifts this critical value of $J$
to the right (figure~\ref{fig:traslado}[b]). For higher excitation numbers,
this region shows higher complexity, whereas the regular areas remain largely
unchanged (see figure~\ref{fig:ensanchmiento}). As concurrence mimics the
behaviour of purity, this observations are also valid for the internal
entanglement. 
A remarkable fact is
also seen: the existence of critical points seems to agree for all cases, for
the density diagrams (figures \ref{fig:traslado} and \ref{fig:ensanchmiento}).
\section{Conclusions and outlook} 
\label{sec:conclusions}
In the present work we studied the entanglement dynamics of three interacting
two level atoms inside a cavity with dipole-dipole and Ising interactions.
Entanglement within the atoms is measured by concurrence, and of the atoms
with the cavity, is measured by purity. Despite measuring 
totally different properties of the system, concurrence and purity are quite 
related. In particular in a concurrence-purity plane for  given state,
concurrence is bounded by above, by the curve described by a non interactive
system, and initially in a W state. We believe that this upper bound is closely
relate with monogamy of entanglement where entanglement cannot be freely shared
among multiple parties~\cite{PhysRevA.61.052306}. Recall that
our system is a four-partite state, namely three qubits and a bath in which 
entanglement must be considered, in this setting, not as a tripartite 
problem. 
At this respect, our analysis contributes to the existence of a
hierarchy of strong monogamy (SM) inequalities (as propose by Regula et
al.~\cite{PhysRevLett.113.110501}), or alternatively viewing multipartite
entanglement from the point of view of
frustration~\cite{1367-2630-12-2-025015}. In the
homogeneous case, a lower bound, drawn by the evolution of the same initial
state, but with a non-interacting Hamiltonian is also apparent. The
inhomogeneous case, however, has not a simple lower bound. In fact, for
this lower bound, calculations could be pursued using the Dicke States and the
symmetric subspace, where the block-diagonal Hamiltonian and the basis,
resembles the two qubit case. In the present work, this bound is very sensitive
to the presence of a small perturbation, so this suggests a strong connection 
with the symmetric properties of the Dicke basis. A deeper study of 
these bound might prove useful in the context of mutipartite entanglement. 
Finally, we presented a global view for the dynamics of the
concurrence and purity as function of time and Ising interaction, and show that
there is a translation of the intervals where the dynamics exhibits complexity,
when the dipole-dipole interaction is increasing. 



\ack 
Support by the  projects CONACyT 153190 and UNAM-PAPIIT IA101713 and
IN111015 is
acknowledged. Numerics were done building upon programs developed
by Florian Minter and André Carvalho, and stimulating discussions with Thomas
Seligman and Mauricio Torres are acknowledged. 
\section*{References}
\bibliographystyle{unsrt}
\bibliography{bibliografia}
\end{document}